\documentclass[utf8]{frontiersSCNS} 
\usepackage{url,hyperref,lineno,microtype,subcaption}
\usepackage[onehalfspacing]{setspace}
\def\keyFont{\fontsize{8}{11}\helveticabold }

\def\firstAuthorLast{Hutsem\'ekers {et~al.}} 
\def\Authors{Damien Hutsem\'ekers\,$^{1,*}$, Lorraine Braibant\,$^{1}$, Dominique Sluse\,$^{1}$, Timo Anguita\,$^{2}$, and René Goosmann\,$^{3}$}
\def\Address{$^{1}$Institut d'Astrophysique et de G\'eophysique, Universit\'e de Li\`ege, All\'ee du six ao\^ut 19c, B-4000 Li\`ege, Belgium\\
             $^{2}$Departamento de Ciencias Fisicas, Universidad Andres Bello, Av. Republica 252, Santiago, Chile\\
             $^{3}$Observatoire Astronomique de Strasbourg, Universit\'e de Strasbourg, Rue de l’Universit\'e 11, F-67000 Strasbourg, France}  
\def\corrAuthor{Damien Hutsem\'ekers}
\def\corrEmail{hutsemekers@astro.ulg.ac.be}

\usepackage{aas_macros}

\begin{document}
\onecolumn
\firstpage{1}

\title[Constraints on quasar BLRs from microlensing]{New constraints on quasar broad absorption and emission line regions from gravitational microlensing} 

\author[\firstAuthorLast ]{\Authors}\address{}\correspondance{}\extraAuth{}
\maketitle

\begin{abstract}
Gravitational microlensing is a powerful tool allowing one to probe the structure of quasars on sub-parsec scale. We report recent results, focusing on the broad absorption and emission line regions. In particular microlensing reveals the intrinsic absorption hidden in the P~Cygni-type line profiles observed in the broad absorption line quasar H1413+117, as well as the existence of an extended continuum source. In addition, polarization microlensing provides constraints on the scattering region. In the quasar Q2237+030, microlensing differently distorts the H$\alpha$ and CIV broad emission line profiles, indicating that the low- and high-ionization broad emission lines must originate from regions with distinct kinematical properties. We also present simulations of the effect of microlensing on line profiles considering simple but representative models of the broad emission line region. Comparison of observations to simulations allows us to conclude that the H$\alpha$ emitting region in Q2237+030 is best represented by a Keplerian disk.

\tiny\keyFont{\section{Keywords:} quasars, broad emission lines, broad absorption lines, gravitational microlensing, polarization} 
\end{abstract}

\section{Introduction}

When the light from a distant quasar passes through the gravitational field of a galaxy, it is deflected and multiple magnified images of the source are observed. In addition, stars in the lensing galaxy can act as microlenses and produce an extra magnification of some images. The collective effect of these stars generates a complex magnification pattern in the source plane that takes the form of a caustic network. Microlensing magnification varies in time due to the relative motions of the source, lens and observer, on timescales of weeks to years. High magnification events occur when caustics are close to the line of sight \citep[e.g.,][for a review]{2010SchmidtWambsganss}.

Gravitational lensing magnification strongly depends on the Einstein radius of the system, in the sense that only sources smaller than this radius can be significantly magnified. For a typical lensed system with a lensing galaxy at redshift $z \sim 0.5$ and a source quasar at $z \sim 2$, the projected Einstein radius of an average 0.3 M$_{\odot}$ star in the lensing galaxy is of the order of 10$^{-3}$ pc, which is comparable to the size of the quasar continuum-emitting accretion disk. Microlensing thus mainly magnifies the continuum source. The more extended broad emission line region is either unaffected or only partly magnified. The much larger narrow line region is totally unaffected.

Therefore, microlensing differently magnifies the various components of the quasar spectrum. By comparing the spectra of two images of a lensed quasar, one affected by microlensing and the other not, one can separate the part of the spectrum which is microlensed, that is the part coming from the most compact source, from the part of the spectrum which is not affected by microlensing and originates from a more extended region.  Information on the geometry and kinematics of the quasar inner regions can thus be obtained. This is illustrated in the following sections, focusing on the broad absorption and emission line regions (respectively, BALR and BELR).

\section{Microlensing in the Broad Absorption Line quasar H1413+117}

H1413+117, the cloverleaf, is a quadruply lensed broad absorption line (BAL) quasar in which a slowly varying microlensing effect lasting for $\sim$ 20 years magnifies the continuum of image D, leaving the emission lines essentially unchanged \citep{1990Angonin,1993Hutsemekers,2008Anguita,2010Hutsemekers,2015Sluse,2015Odowd}. Image A is not or weakly affected by microlensing, thus providing a reference spectrum. We can then separate the part of the line profiles that is microlensed from the part that is not microlensed using the macro-micro decomposition (MmD) method presented in \citet{2007Sluse} and \citet{2010Hutsemekers}.

\begin{figure}[h!]
\begin{center}
\includegraphics[height=8.5cm]{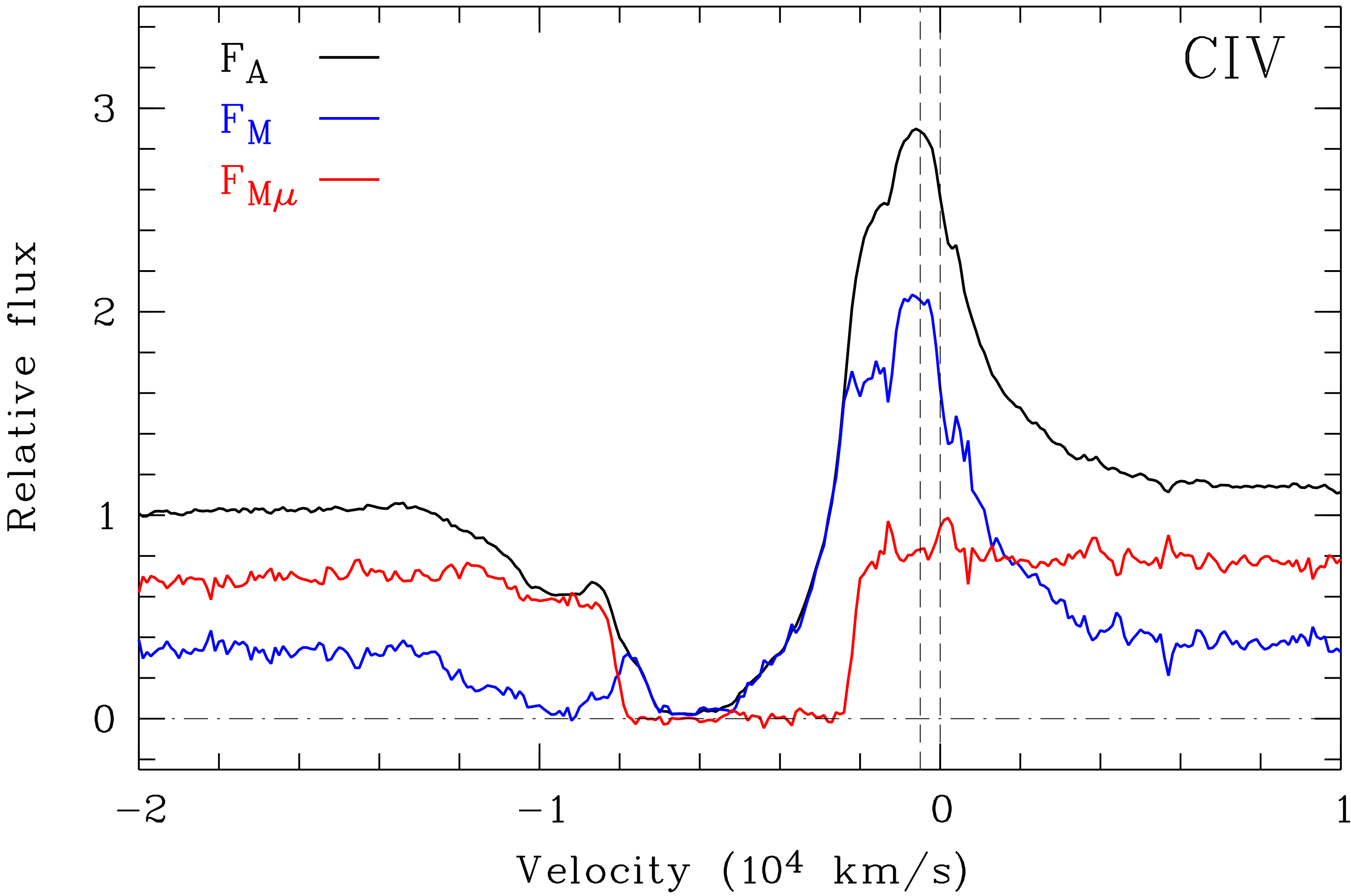}
\end{center}
\caption{Microlensing in the CIV line of the BAL quasar H1413+117. $F_A$ is the observed spectrum of image A which is not affected by microlensing. The line profile can be decomposed in a part affected by microlensing, $F_{M\mu}$, and another part not affected by microlensing, $F_M$, with $F_A = F_M + F_{M\mu}$.}
\label{fig:1}
\end{figure}

Fig.~\ref{fig:1} illustrates the decomposition of the CIV~$\lambda$1549\AA\ P~Cygni-type line profile into its microlensed (F$_{M\mu}$) and non-microlensed (F$_{M}$) components \citep[see][for more details]{2010Hutsemekers,2015Sluse}. F$_{M\mu}$ comes from a compact region that essentially produces the absorbed continuum, while F$_{M}$ comes from a more extended region too large to be microlensed. From this spectral decomposition, it results that (1) the BAL profile consists of a completely black absorption (seen in F$_{M\mu}$) partially filled in by the broad emission line (seen in F$_{M}$); (2) this absorption does not start at zero-velocity but at an onset velocity of $\sim$ 2000 km/s; (3) the broad emission line itself is re-absorbed (see F$_{M}$) over a wavelength range narrower than the full absorption profile, revealing the existence of an additional, more extended BALR; (4) a part of the continuum (seen in F$_{M}$) is not microlensed thus originating from a region larger than the source of the continuum seen in F$_{M\mu}$. These observations suggest a two-component outflow: one component is co-spatial with the BELR, while the other one, more distant, partially re-absorbs the emission from the BELR \citep{2010Borguet}. \citet{2015Odowd} found in addition an ionization dependence of the BAL onset velocity that they interpreted in the framework of a disk-wind model.

The light from H1413+117 is linearly polarized and \citet{2001Chae} found  evidence for polarization microlensing. \citet{2015Hutsemekers} obtained spectropolarimetric observations of the different images that suggest that the continuum is scattered off two regions generating roughly orthogonal polarizations: a compact region located in the equatorial plane close to the accretion disk, which is microlensed, and an extended region located along the polar axis, which is not microlensed. This indicates that the non-microlensed extended continuum seen in F$_{M}$ (Fig.~\ref{fig:1}) could originate from scattering.

\section{Microlensing of the broad emission line region in Q2237+030}

\subsection{Observed line profile distortions}

Though larger than the source of continuum, the BELR can be partly magnified by microlensing, which results in line profile deformations.  Such microlensing-induced line profile deformations have been observed in several lensed quasar spectra, exhibiting various symmetric and asymmetric distortions in both low- and high-ionization lines \citep{2004Richards, 2005Wayth, 2007Sluse, 2011Sluse, 2012Sluse, 2011ODowd, 2013Guerras, 2014Braibant, 2016Braibant, 2016Goicoechea, 2017Motta}. By selectively magnifying different subregions of the broad line region, gravitational microlensing can thus provide information on the size, geometry and kinematics of the BELR in quasars \citep{1988Nemiroff, 1990SchneiderWambsganss}. In particular, the size of the BELR has been estimated in a few objects and found compatible with reverberation mapping measurements \citep{2005Wayth, 2011Sluse, 2013Guerras}.

\begin{figure}[h!]
\begin{center}
\includegraphics[height=8.5cm]{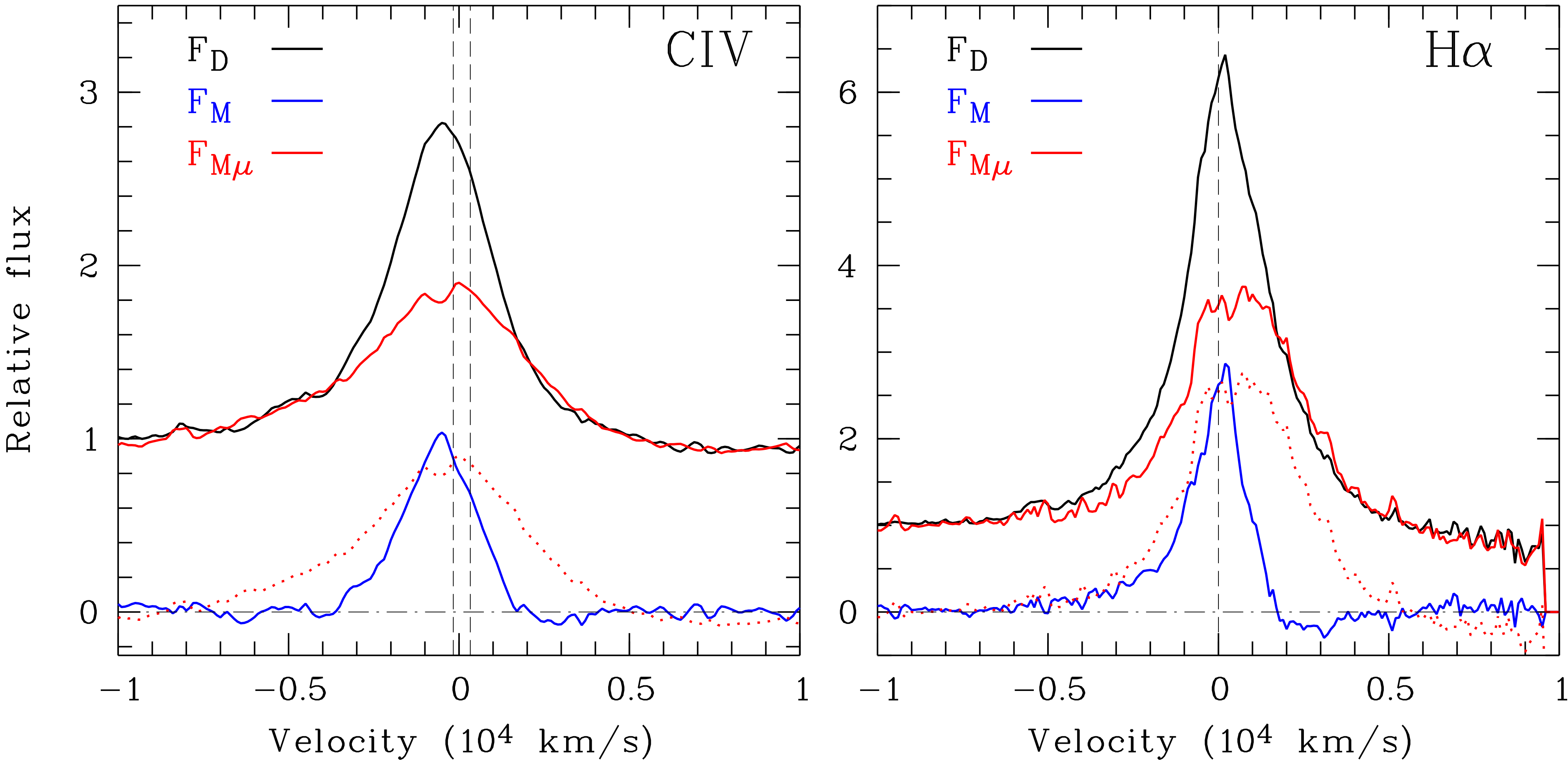}
\end{center}
\caption{Microlensing in the CIV and H$\alpha$ lines of the quasar Q2237+030. $F_D$ is the observed spectrum of image D which is not affected by microlensing. The line profile can be decomposed in a part affected by microlensing, $F_{M\mu}$, and another part not affected by microlensing, $F_M$, with $F_D = F_M + F_{M\mu}$. The dotted line profile represents the continuum-subtracted $F_{M\mu}$.}
\label{fig:2}
\end{figure}

Making use of the MmD line profile disentangling technique, \cite{2014Braibant, 2016Braibant} have extracted the part of the emission line profiles that is affected by microlensing in two lensed quasars for which high quality data were available. Fig.~\ref{fig:2} illustrates the F$_{M\mu}$ $/$ F$_{M}$ decomposition of the CIV~$\lambda$1549\AA\ and H$\alpha$ broad emission line profiles for the quadruply lensed quasar Q2237+030, the Einstein Cross, at an epoch when microlensing was prominent in image A and negligible in image D. A clear difference in the distortions suffered by the CIV and H$\alpha$ emission lines can be noticed. In CIV the effect is essentially symmetric, the wings of the line being more magnified than the core, while in H$\alpha$ the effect is asymmetric, the red wing being more magnified than the blue wing. Since these observations were obtained at the same epoch, both the CIV and H$\alpha$ emitting regions are magnified by the same caustic pattern. This indicates that the high- and low-ionization emission lines must originate from regions with distinct kinematical properties.

\begin{figure}[h!]
\begin{center}
\includegraphics[width=\hsize]{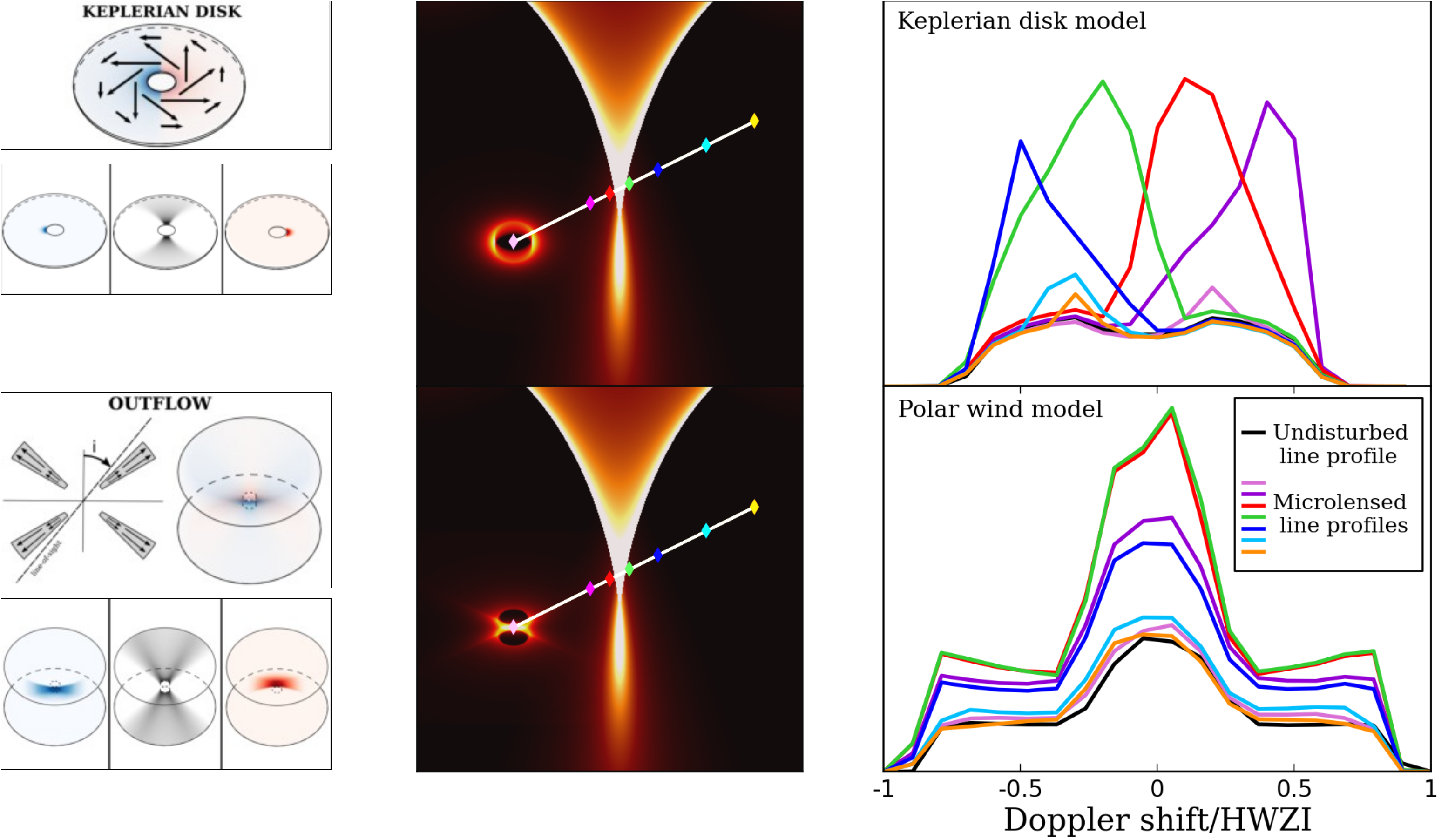}
\end{center}
\caption{Microlensing of different BELR structures and the resulting line profile distortions. The left panel sketches two models: a rotating Keplerian disk and a biconical outflow, both seen at intermediate inclination. The location of the approaching and receding gas is indicated in blue and red, respectively. The middle panel shows the BELR superimposed on a typical caustic. At the different positions indicated by colored losanges, different subregions of the BELR are magnified by the caustic. The right panel illustrates the line profiles corresponding to the different positions of the BELR onto the caustic pattern. Microlensing of the Keplerian disk is characterized by asymmetric red/blue line profile distortions while microlensing of the biconical outflow is characterized by symmetric wings/core distortions.}
\label{fig:3}
\end{figure}

As illustrated in Fig.~\ref{fig:3}, the detection of a red/blue differential microlensing effect in the H$\alpha$ line profile suggests that the low-ionization BELR is likely a Keplerian disk, while the wings/core distortion observed in the CIV line can be interpreted assuming a polar outflow model. However, the mapping between a wavelength range in the line profile and subregions of the BELR seen in projection is usually not unique and a confrontation of the observations to detailed modeling appeared necessary.

\subsection{Simulations of line profile distortions}

Possible effects of microlensing on broad emission lines have been theoretically investigated by several authors \citep{1988Nemiroff, 1990SchneiderWambsganss, 1994Hutsemekers, 2001Popovic, 2002Abajas, 2007Abajas, 2004LewisIbata, 2011ODowd, 2014Simic} considering various BELR models and magnification patterns. 

In \citet{2017Braibant}, we extend those works focusing on wings/core and red/blue line profile distortions in order to constrain the BELR models. The effects of gravitational microlensing on the quasar broad emission line profiles and their underlying continuum have been simultaneously computed, considering simple, representative BELR, accretion disk, and magnifying caustic models. Keplerian disks as well as polar and equatorial outflow models of various sizes have been considered. The effect of microlensing has been quantified using four observables: $\mu^{BLR}$, the total magnification of the broad emission line, $\mu^{cont}$, the magnification of the underlying continuum, as well as red/blue, $RBI$, and wings/core, $WCI$, indices that characterize the line profile distortions. Those quantities were designed to not depend on the exact profile of the broad emission lines, so that they can be directly compared to observations.

The simulations show that asymmetric distortions of broad line profiles like those reported in \cite{2014Braibant, 2016Braibant} can be reproduced, and attributed to the differential effect of microlensing on spatially and kinematically separated regions of the BELR. In particular, red/blue asymmetric distortions constitute a good discriminant between the polar outflow and other models. We then built diagrams that can serve as diagnostic tools to discriminate between the different BELR models making use of quantitative measurements of the four observables $\mu^{cont}$, $\mu^{BLR}$, $RBI$, $WCI$. It appeared from the simulations that only strong microlensing effects can produce distinctive line profile distortions for a limited number of caustic configurations, i.e., distortions that allow us to put serious constraints on the various BELR models.

\begin{figure}[h!]
\begin{center}
\includegraphics[height=5.5cm]{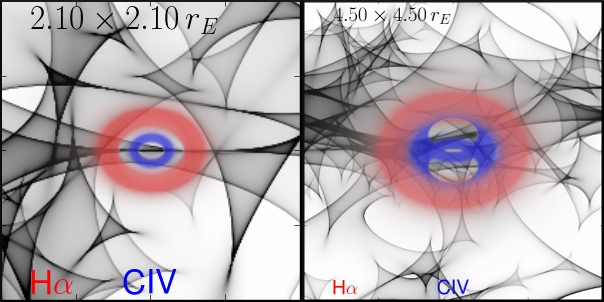}
\end{center}
\caption{Examples of models of the CIV and H$\alpha$ BELRs and their location on the caustic network that simultaneously fit the microlensing observables $\mu^{cont}$, $\mu^{BLR}$, $RBI$,  $WCI$ measured for both the CIV and H$\alpha$ lines in the lensed quasar Q2237+030. Left: Keplerian disks for both H$\alpha$ and CIV BELRs. Right: Keplerian disk for H$\alpha$ and polar wind for CIV. All models are seen under intermediate inclination, $i = 44^{\circ}$. The CIV emitting region (in blue) is more compact than the H$\alpha$ emitting region (in red).}
\label{fig:4}
\end{figure}

The four indices $\mu^{cont}$, $\mu^{BLR}$, $RBI$, $WCI$ were then measured for both the CIV and H$\alpha$ emission lines observed in the lensed quasar Q2237+030, and compared to the values generated by the simulations (Braibant et al., in preparation). In the simulations, we assumed that the CIV and H$\alpha$ BELRs, and the continuum-emitting accretion disk share the same inclination with respect to the line of sight and the same location on the caustic network. From this comparison it results that the H$\alpha$ low-ionization BELR is best represented by a Keplerian disk while the CIV high-ionization BELR can be either a Keplerian disk or a polar wind. Examples of fitting models are illustrated in Fig.~\ref{fig:4}. In all cases the H$\alpha$ BELR is roughly a factor 2 larger than the CIV BELR. Recent results from velocity-resolved reverberation mapping also suggest that the low-ionization BELR can be a Keplerian disk \citep[][and references therein]{2017Grier}.

\section{Conclusions}

These results demonstrate the potential of microlensing to probe the geometry and kinematics of the broad line regions and outflows in quasars. Constraints on the location of the scattering regions at the origin of the polarization can also be obtained. Simulations of line profile distortions show that only strong microlensing effects can produce line profile distortions allowing us to discriminate between various BELR models using single epoch data. To benefit from weaker microlensing effects, statistical analyses of larger data sets would be needed. In particular, a long-term spectrophotometric monitoring of the different images of lensed quasars would provide a real scan of their line emitting regions, with the possibility to constrain more complex models than those considered here. Microlensing can thus be a powerful and alternative approach to reverberation mapping, especially since it can be applied to high redshift quasars, with little dependence on their luminosity, and to the study of both the low- and high-ionization BLRs.

\section*{Conflict of Interest Statement}
The authors declare that the research was conducted in the absence of any commercial or financial relationships that could be construed as a potential conflict of interest.

\section*{Author Contributions}
DH and DS initiated the project. LB, DH, DS, and TA contributed to the observations, data reduction and analysis. LB, DH, DS and RG contributed to the BELR microlensing modeling. All authors contributed to the discussions. DH wrote the paper with contributions from DS. LB made the artwork in figure~\ref{fig:3}.

\section*{Acknowledgments}
DH and LB acknowledge support of Belgian F.R.S.-FNRS. Support for TA is provided by project FONDECYT 11130630 and the Ministry of Economy, Development, and Tourism’s Millennium Science Initiative through grant IC120009, awarded to The Millennium Institute of Astrophysics, MAS.

\bibliographystyle{frontiersinSCNS_ENG_HUMS}
\bibliography{hutsemekers}

\section*{Figure captions}

\end{document}